\documentclass[english]{article}
\usepackage{amsfonts,amsmath,amsthm,amscd,amssymb,latexsym}
\usepackage[active]{srcltx}                    

\theoremstyle{plain}
\newtheorem{thm}{Theorem}[section]
\newtheorem{cor}[thm]{Corollary}

\newtheorem{prop}[thm]{Proposition}

\theoremstyle{definition}
\theoremstyle{remark}
\numberwithin{equation}{section}
\newcommand{\keywords}{\textbf{Key words and phrases: }\medskip}
\newcommand{\subjclass}{\textbf{Math. Subj. Clas.: }\medskip}

\DeclareMathOperator{\RE}{Re} \DeclareMathOperator{\IM}{Im}

\begin{document}
\title{\textbf{Discrete Dirac-K\"{a}hler equation and its formulation in algebraic form} }
\author{\textbf{Volodymyr Sushch} \\
{ Koszalin University of Technology} \\
 { Sniadeckich 2, 75-453 Koszalin, Poland} \\
 {volodymyr.sushch@tu.koszalin.pl} }

\date{}
\maketitle
\begin{abstract}
A relationship between the discrete Dirac-K\"{a}hler  equation  and  discrete analogues of some  Dirac type equations  in the geometric spacetime  algebra is discussed. We show that a solution of the discrete Dirac-K\"{a}hler equation can be represented  as the sum and difference  of solutions of   discrete Dirac type equations with the corresponding  sign of the mass term.
\end{abstract}

\bigskip
\medskip


\section{Introduction}
This work is a direct continuation of my previous paper \cite{S3} in which a correspondence between the discrete   Dirac-K\"{a}hler  equation and a discrete analogue of the Hestenes equation is studied.
In \cite{S3}, it was shown that the geometric discretization scheme as developed in \cite{S2} can be used to find a new discrete formulation of the Dirac equation for a free electron in the Hestenes form.
The purpose of this  paper is to discuss a relationship between the discrete  Hestenes equation  and some other discrete Dirac type equations which are formulated  in the algebraic form.

We first briefly review some definitions and basic notation on the Dirac-K\"{a}hler equation \cite{Kahler, Rabin}.
Let $M={\mathbb R}^{1,3}$ be  Minkowski space with  metric signature  $(+,-,-,-)$.
Denote by $\Lambda^r(M)$ the vector space of smooth differential $r$-forms, $r=0,1,2,3,4$. We consider  $\Lambda^r(M)$ over $\mathbb{C}$.
Let $d:\Lambda^r(M)\rightarrow\Lambda^{r+1}(M)$ be the exterior differential and let $\delta:\Lambda^r(M)\rightarrow\Lambda^{r-1}(M)$ be the formal adjoint of $d$  with respect to  the natural inner product in $\Lambda^r(M)$ (codifferential). We have $\delta=\ast d\ast$, where $\ast:\Lambda^r(M)\rightarrow\Lambda^{4-r}(M)$ is the Hodge star operator  with respect to the Lorentz metric.
 Denote by $\Lambda(M)$ the set of all differential forms on $M$. We have
\begin{equation*}
\Lambda(M)=\Lambda^0(M)\oplus\Lambda^1(M)\oplus\Lambda^2(M)\oplus\Lambda^3(M)\oplus\Lambda^4(M)=\Lambda^{ev}(M)\oplus\Lambda^{od}(M),
\end{equation*}
where $\Lambda^{ev}(M)=\Lambda^0(M)\oplus\Lambda^2(M)\oplus\Lambda^4(M)$  and  $\Lambda^{od}(M)=\Lambda^1(M)\oplus\Lambda^3(M)$.
Let $\Omega\in\Lambda(M)$
be an inhomogeneous differential form, i.e.
\begin{equation*}\Omega=\sum_{r=0}^4\overset{r}{\omega}, \quad \overset{r}{\omega}\in\Lambda^r(M).
\end{equation*}
 The Dirac-K\"{a}hler equation is given by
\begin{equation}\label{1.1}
i(d+\delta)\Omega=m\Omega,
\end{equation}
where $i$ is the usual complex unit ($i^2=-1$) and  $m$  is a mass parameter.
It is easy to show that Equation~(\ref{1.1}) is equivalent to the set of equations
\begin{eqnarray*}\label{}
i\delta\overset{1}{\omega}=m\overset{0}{\omega},\\
i(d\overset{0}{\omega}+\delta\overset{2}{\omega})=m\overset{1}{\omega},\\
i(d\overset{1}{\omega}+\delta\overset{3}{\omega})=m\overset{2}{\omega},\\
i(d\overset{2}{\omega}+\delta\overset{4}{\omega})=m\overset{3}{\omega},\\
id\overset{3}{\omega}=m\overset{4}{\omega}.
\end{eqnarray*}
The operator $d+\delta$ is an analogue of the gradient operator in Minkowski spacetime
\begin{equation*}
\nabla=\sum_{\mu=0}^3\gamma_\mu\partial^\mu, \quad \mu=0,1,2,3,
\end{equation*}
 where $\gamma_\mu$ is the Dirac gamma matrix. Think of $\{\gamma_0, \gamma_1, \gamma_2, \gamma_3\}$ as a vector basis in spacetime. Then the  gamma matrices $\gamma_\mu$ can be considered as generators of the Clifford algebra of spacetime $\emph{C}\ell(1,3)$ (the spacetime algebra) \cite{B, H2}.
In the algebraic formulation the explicit matrices that represent the basis $\{\gamma_\mu\}$ are not important.
Denote by $\emph{C}\ell_{\mathbb{R}}(1,3)$ ($\emph{C}\ell_{\mathbb{C}}(1,3)$) the real (complex) Clifford algebra.  It is known that an inhomogeneous form $\Omega$ can be represented as element of $\emph{C}\ell_{\mathbb{C}}(1,3)$. Then the Dirac-K\"{a}hler equation can be written as an algebraic equation
\begin{equation}\label{1.2}
 i\nabla\Omega=m\Omega, \quad \Omega\in\emph{C}\ell_{\mathbb{C}}(1,3).
 \end{equation}
  Equation~(\ref{1.2}) is equivalent to the four Dirac equations (traditional column-spinor equations) for a free electron.  Let $\emph{C}\ell^{ev}(1,3)$ be the even subalgebra of  $\emph{C}\ell(1,3)$.  Then the following equation
 \begin{equation}\label{1.3}
 -\nabla\Omega\gamma_1\gamma_2=m\Omega\gamma_0, \quad \Omega\in\emph{C}\ell_{\mathbb{R}}^{ev}(1,3),
 \end{equation}
is called the Hestenes form of the Dirac equation \cite{H1, H2}. The Hestenes equation is equivalent to the conventional Dirac equation \cite{H1, Marchuk}.
We consider two more Dirac type equations in the algebraic formulation, namely
\begin{equation}\label{1.4}
 i\nabla\Omega=m\Omega\gamma_0 , \quad \Omega\in\emph{C}\ell_{\mathbb{C}}^{ev}(1,3)
 \end{equation}
 and
 \begin{equation}\label{1.5}
 -\nabla\Omega=m\Omega\gamma_0\gamma_1\gamma_2\gamma_3 , \quad \Omega\in\emph{C}\ell_{\mathbb{R}}(1,3).
 \end{equation}
 Equation~(\ref{1.4}) is described in  \cite{J} and this is the so-called  generalized bivector Dirac equation. Baylis \cite{B} calls Equation~(\ref{1.4}) the Joyce equation. In \cite{Marchuk}, Equation~(\ref{1.5}) is derived from the Dirac-K\"{a}hler equation.
It should be noted  that the algebra $\Lambda(M)$ endowed with the Clifford multiplication is an example of the the Clifford algebra.
In this case the basis covectors $dx^\mu$, $\mu=0,1,2,3$, of spacetime are considered as generators of this algebra. Hence, taking $dx^\mu$ instead of $\gamma_\mu$  Equations~(\ref{1.2})--(\ref{1.4})  can be rewritten in terms of inhomogeneous forms.

 In this paper we construct  discrete analogues  of  Equations~(\ref{1.4}) and (\ref{1.5}). In \cite{B}, a correspondence between the Dirac equation in  the Hestenes form and the Joyce equation is established. We describe a straightforward discrete version of this correspondence by using the technique of projectors which act in  the space of discrete forms.
In much the same way as in the continuum case \cite{B}  a solution of discrete Equation~(\ref{1.4}) or (\ref{1.5}) can be decomposed into solutions of the discrete
Dirac-K\"{a}hler  equation or the discrete Hestenes equation.
We show that every solution of the discrete Dirac-K\"{a}hler equation can be expressed as a linear combination of real-valued solutions of the discrete Equation~(\ref{1.5}) with the correct or reversed sign of the mass term.

\section{Discrete Dirac-K\"{a}hler equation}
Our starting point is a  discretization scheme for the Dirac-K\"{a}hler equation. We use the geometric scheme   based on the language of differential forms in which the differential $d$ and codifferential $\delta$ are replaced by their discrete analogues. This construction is adapted from   \cite{S2}.
For the convenience of the reader we repeat the relevant material from  \cite{S2}, thus making our presentation self-contained. We refer the reader to  \cite{S1, S2} for full mathematical details of the approach. Note that this approach was originated by Dezin \cite{Dezin}.

Let $K(4)=K\otimes K\otimes K\otimes K$
be a cochain complex with  complex  coefficients,
where  $K$ is  the 1-dimensional complex generated by 0- and 1-dimensional basis elements   $x^{\kappa}$  and $e^{\kappa}$,  $\kappa\in\mathbb{Z}$,  respectively.
Then an arbitrary r-dimensional basis element of $K(4)$ can be written as
$s^k_{(r)}=s^{k_0}\otimes s^{k_1}\otimes s^{k_2}\otimes s^{k_3}$, where
$s^{k_\mu}$ is either $x^{k_\mu}$ or $e^{k_\mu}$,  $k=(k_0,k_1,k_2,k_3)$ and \ $k_\mu\in\mathbb{Z}$.
The dimension $r$ of a basis element $s^k_{(r)}$ is given
by the number of factors $e^{k_\mu}$ that appear in it. For example, the 1-dimensional basis elements
of $K(4)$ can be written as
\begin{eqnarray*}
e^k_0=e^{k_0}\otimes x^{k_1}\otimes x^{k_2}\otimes x^{k_3},  \qquad
e^k_1=x^{k_0}\otimes e^{k_1}\otimes x^{k_2}\otimes x^{k_3}, \\
e^k_2=x^{k_0}\otimes x^{k_1}\otimes e^{k_2}\otimes x^{k_3},  \qquad
e^k_3=x^{k_0}\otimes x^{k_1}\otimes x^{k_2}\otimes e^{k_3},
\end{eqnarray*}
where  the subscript $\mu=0,1,2,3$ indicates  a place of $e^{k_\mu}$ in $e^k_\mu$. Similarly,   $e_{\mu\nu}^k$, $\mu<\nu$, and  $e_{\iota\mu\nu}^k$, $\iota<\mu<\nu$, denote the   2- and 3-dimensional basic elements of $K(4)$.
The complex $K(4)$ is a discrete analogue of $\Lambda(M)$ and cochains play the role of differential
forms. Let us call  them forms or discrete forms to emphasize their relationship with differential
forms.

 Denote by  $K^r(4)$ the set of all $r$-forms. Then we have
\begin{equation*}
K(4)=K^0(4)\oplus K^1(4)\oplus K^2(4)\oplus K^3(4)\oplus K^4(4)=K^{ev}(4)\oplus K^{od}(4),
\end{equation*}
 where $K^{ev}(4)=K^0(4)\oplus K^2(4)\oplus K^4(4)$ and $K^{od}(4)=K^1(4)\oplus K^3(4)$.
We can expand an  $r$-form $\overset{r}{\omega}\in K^r(4)$ as
 \begin{equation}\label{2.1}
\overset{r}{\omega}=\sum_k\sum_r\omega_k^{(r)}s^k_{(r)}, \qquad  \omega_k^{(r)}\in\mathbb{C}.
\end{equation}
 More precisely,  we have
\begin{equation}\label{2.2}
\overset{0}{\omega}=\sum_k\overset{0}{\omega}_kx^k,  \qquad  x^k=x^{k_0}\otimes x^{k_1}\otimes x^{k_2}\otimes x^{k_3},
\end{equation}
\begin{equation}\label{2.3}
 \overset{4}{\omega}=\sum_k\overset{4}{\omega}_ke^k, \qquad  e^k=e^{k_0}\otimes e^{k_1}\otimes e^{k_2}\otimes e^{k_3},
\end{equation}
\begin{eqnarray}\label{2.4}
\overset{1}{\omega}=\sum_k\sum_{\mu=0}^3\omega_k^\mu e_\mu^k, \quad
\overset{2}{\omega}=\sum_k\sum_{\mu<\nu} \omega_k^{\mu\nu}e_{\mu\nu}^k, \quad
\overset{3}{\omega}=\sum_k\sum_{\iota<\mu<\nu} \omega_k^{\iota\mu\nu}e_{\iota\mu\nu}^k,
\end{eqnarray}
where the components $\overset{0}{\omega}_k, \ \overset{4}{\omega}_k, \  \omega_k^\mu, \ \omega_k^{\mu\nu}$ and $\omega_k^{\iota\mu\nu}$ are complex numbers.
A discrete inhomogeneous form $\Omega\in K(4)$  is defined to be
\begin{equation}\label{2.5}
\Omega=\sum_{r=0}^4\overset{r}{\omega},
\end{equation}
where $\overset{r}{\omega}$ is given by (\ref{2.1}).

Let $d^c: K^r(4)\rightarrow K^{r+1}(4)$ be a discrete analogue of the exterior derivative $d$ and let $\delta ^c: K^r(4)\rightarrow K^{r-1}(4)$ be a discrete analogue of the codifferential $\delta$. For definitions of these operators  we refer the reader to  \cite{S2}. In this paper  we give only the difference
expression for $d^c$ and  $\delta ^c$.

Let the difference operator $\Delta_\mu$ be defined by
\begin{equation}\label{2.6}
\Delta_\mu\omega_k^{(r)}=\omega_{\tau_\mu k}^{(r)}-\omega_k^{(r)},
\end{equation}
where
$\tau_\mu$ is   the shift operator  which acts  as
\begin{equation*}
\tau_\mu k=(k_0,...k_\mu+1,...k_3), \quad   \mu=0,1,2,3.
\end{equation*}
For forms (\ref{2.2})--(\ref{2.4})  we have
\begin{eqnarray}\label{2.7}
d^c\overset{0}{\omega}=\sum_k\sum_{\mu=0}^3(\Delta_\mu\overset{0}{\omega}_k)e_\mu^k,  \qquad d^c\overset{1}{\omega}=\sum_k\sum_{\mu<\nu}(\Delta_\mu\omega_k^\nu-\Delta_\nu\omega_k^\mu)e_{\mu\nu}^k,
\end{eqnarray}
\begin{eqnarray}\label{2.8}
d^c\overset{2}{\omega}=\sum_k\big((\Delta_0\omega_k^{12}-\Delta_1\omega_k^{02}+\Delta_2\omega_k^{01})e_{012}^k \\ \nonumber
+(\Delta_0\omega_k^{13}-\Delta_1\omega_k^{03}+\Delta_3\omega_k^{01})e_{013}^k \\  \nonumber
+(\Delta_0\omega_k^{23}-\Delta_2\omega_k^{03}+\Delta_3\omega_k^{02})e_{023}^k \\ \nonumber
+(\Delta_1\omega_k^{23}-\Delta_2\omega_k^{13}+\Delta_3\omega_k^{12})e_{123}^k\big),
\end{eqnarray}
\begin{equation}\label{2.9}
d^c\overset{3}{\omega}=\sum_k(\Delta_0\omega_k^{123}-\Delta_1\omega_k^{023}+\Delta_2\omega_k^{013}-\Delta_3\omega_k^{012})e^k, \qquad d^c\overset{4}{\omega}=0,
\end{equation}
\begin{equation}\label{2.10}
\delta^c\overset{0}{\omega}=0, \qquad \delta^c\overset{1}{\omega}=\sum_k(\Delta_0\omega_k^{0}-\Delta_1\omega_k^{1}-\Delta_2\omega_k^{2}-\Delta_3\omega_k^{3})x^k,
\end{equation}
\begin{eqnarray}\label{2.11}
\delta^c\overset{2}{\omega}=\sum_k\big((\Delta_1\omega_k^{01}+\Delta_2\omega_k^{02}+\Delta_3\omega_k^{03})e_{0}^k \\ \nonumber
+(\Delta_0\omega_k^{01}+\Delta_2\omega_k^{12}+\Delta_3\omega_k^{13})e_{1}^k\\ \nonumber
+(\Delta_0\omega_k^{02}-\Delta_1\omega_k^{12}+\Delta_3\omega_k^{23})e_{2}^k \\ \nonumber
+(\Delta_0\omega_k^{03}-\Delta_1\omega_k^{13}-\Delta_2\omega_k^{23})e_{3}^k\big),
\end{eqnarray}
\begin{eqnarray}\label{2.12}
\delta^c\overset{3}{\omega}=\sum_k\big((-\Delta_2\omega_k^{012}-\Delta_3\omega_k^{013})e_{01}^k+
(\Delta_1\omega_k^{012}-\Delta_3\omega_k^{023})e_{02}^k\\ \nonumber
+(\Delta_1\omega_k^{013}+\Delta_2\omega_k^{023})e_{03}^k
+(\Delta_0\omega_k^{012}-\Delta_3\omega_k^{123})e_{12}^k\\ \nonumber
+(\Delta_0\omega_k^{013}+\Delta_2\omega_k^{123})e_{13}^k
+(\Delta_0\omega_k^{023}-\Delta_1\omega_k^{123})e_{23}^k\big),
\end{eqnarray}
\begin{eqnarray}\label{2.13}
\delta^c\overset{4}{\omega}=\sum_k\big((\Delta_3\overset{4}{\omega}_k)e_{012}^k-(\Delta_2\overset{4}{\omega}_k)e_{013}^k
+(\Delta_1\overset{4}{\omega}_k)e_{023}^k+(\Delta_0\overset{4}{\omega}_k)e_{123}^k\big).
\end{eqnarray}
Let $\Omega\in K(4)$ be given by (\ref{2.5}). A discrete analogue of the Dirac-K\"{a}hler equation~(\ref{1.1}) can be defined as
 \begin{equation}\label{2.14}
i(d^c+\delta^c)\Omega=m\Omega.
\end{equation}
We can write this equation more explicitly by separating its homogeneous components as
\begin{eqnarray}\label{2.15}
i\delta^c\overset{1}{\omega}=m\overset{0}{\omega}, \\ \nonumber
i(d^c\overset{0}{\omega}+\delta^c\overset{2}{\omega})=m\overset{1}{\omega}, \\ \nonumber
i(d^c\overset{1}{\omega}+\delta^c\overset{3}{\omega})=m\overset{2}{\omega}, \\ \nonumber
i(d^c\overset{2}{\omega}+\delta^c\overset{4}{\omega})=m\overset{3}{\omega}, \\ \nonumber
id^c\overset{3}{\omega}=m\overset{4}{\omega}.
\end{eqnarray}
Substituting (\ref{2.7})--(\ref{2.13})  into (\ref{2.15}) yields the set of 16 difference equations   \cite{S2}.

\section{Algebraic form of discrete Dirac type  equations}
As in \cite{S3},  we define  the Clifford multiplication of the basis elements $x^k$ and $e^k_\mu$,  \  $\mu=0,1,2,3$, by the following rules:

\medskip
(a) $x^kx^k=x^k, \quad x^ke^k_\mu=e^k_\mu x^k=e^k_\mu$;

(b) $e^k_\mu e^k_\nu+e^k_\nu e^k_\mu=2g_{\mu\nu}x^k$;

(c) $e^k_{\mu_1}\cdots e^k_{\mu_s}=e^k_{\mu_1\cdots \mu_s}$ \ for \ $0\leq \mu_1<\cdots <\mu_s\leq 3$.
\medskip

Here  $g_{\mu\nu}=diag(1,-1,-1,-1)$ is the metric tensor.
Note that the multiplication is defined for the basis elements of $K(4)$ with the same multi-index $k=(k_0,k_1,k_2,k_3)$ supposing the product to be zero in all other cases.
The operation is linearly extended to arbitrary discrete forms. For example, for any  $\overset{1}{\omega}\in K^1(4)$ and  $\overset{2}{\omega}\in K^2(4)$ given by (\ref{2.4}) we have
\begin{eqnarray*}
\overset{1}{\omega}\overset{2}{\omega}=\Big(\sum_k\sum_{\mu=0}^3\omega^\mu_ke_\mu^k\Big)\Big(\sum_k\sum_{\mu<\nu} \omega_k^{\mu\nu}e_{\mu\nu}^k\Big)\\=
\sum_k \big( (\omega^1_k\omega^{01}_k+\omega^2_k\omega^{02}_k+\omega^3_k\omega^{03}_k)e_0^k+
(\omega^0_k\omega^{01}_k+\omega^2_k\omega^{12}_k+\omega^3_k\omega^{13}_k)e_1^k \\ +
(\omega^0_k\omega^{02}_k-\omega^1_k\omega^{12}_k+\omega^3_k\omega^{23}_k)e_2^k  +
(\omega^0_k\omega^{03}_k-\omega^1_k\omega^{13}_k-\omega^2_k\omega^{23}_k)e_3^k \big) \\
+\sum_k \big((\omega^0_k\omega^{12}_k-\omega^1_k\omega^{02}_k+\omega^2_k\omega^{01}_k)e_{012}^k+
(\omega^0_k\omega^{13}_k-\omega^1_k\omega^{03}_k+\omega^3_k\omega^{01}_k)e_{013}^k \\ +
(\omega^0_k\omega^{23}_k-\omega^2_k\omega^{03}_k+\omega^3_k\omega^{02}_k)e_{023}^k
+(\omega^1_k\omega^{23}_k-\omega^2_k\omega^{13}_k+\omega^3_k\omega^{12}_k)e_{123}^k\big).
\end{eqnarray*}
Consequently, the form $\overset{1}{\omega}\overset{2}{\omega}$ is inhomogeneous and $\overset{1}{\omega}\overset{2}{\omega}\in K^{od}(4)$.
\bigskip
\begin{prop}
Let
\begin{equation}\label{3.1}
x=\sum_kx^k, \qquad e_\mu=\sum_ke_\mu^k, \qquad \mu=0,1,2,3.
\end{equation}
Then
\begin{equation}\label{3.2}
e_\mu e_\nu+e_\nu e_\mu=2g_{\mu\nu}x, \qquad \nu=0,1,2,3.
\end{equation}
\end{prop}
\bigskip
\begin{proof}
By the rule (b) it is obvious.
\end{proof}

 Note that the unit 0-form $x$ plays  a role of the unit element in $K(4)$, i.e. for any $r$-form  $\overset{r}{\omega}$ we have $x\overset{r}{\omega}=\overset{r}{\omega}x=\overset{r}{\omega}$. We denote the unit 4-form by
 \begin{equation}\label{3.3}
e=\sum_ke^k,  \qquad e^k=e^{k_0}\otimes e^{k_1}\otimes e^{k_2}\otimes e^{k_3}.
\end{equation}
  It is clear that
\begin{equation}\label{3.4}
e=e_0e_1e_2e_3 \quad \mbox{and} \quad e^2=ee=-x.
\end{equation}

Let us introduce  the following constant forms
\begin{equation}\label{3.5}
P_{\pm 0}=\frac{1}{2}(x\pm e_0), \qquad  P_{\pm 12}=\frac{1}{2}(x\pm ie_1e_2), \qquad  P_{\pm e}=\frac{1}{2}(x\pm ie).
\end{equation}
   It is easy to check that
 \begin{equation*}
(P_{\pm 0})^2=P_{\pm 0}P_{\pm 0}=P_{\pm 0}, \qquad  (P_{\pm 12})^2=P_{\pm 12}P_{\pm 12}=P_{\pm 12},
\end{equation*}
\begin{equation*}
 (P_{\pm e})^2=P_{\pm e}P_{\pm e}=P_{\pm e}.
\end{equation*}
 Hence, the forms $P_{\pm 0}$, $P_{\pm 12}$ and $P_{\pm e}$ are projectors.
\bigskip
\begin{prop}
The projectors $P_{\pm 0}$, $P_{\pm 12}$ and $P_{\pm e}$ have the following properties
\begin{equation}\label{3.6}
P_{\pm 0}P_{\pm 12}=P_{\pm 12}P_{\pm 0}, \quad P_{\pm e}P_{\pm 12}=P_{\pm 12}P_{\pm e},
\end{equation}
\begin{equation}\label{3.7}
 e_0P_{\pm 0}=P_{\pm 0}e_0,  \quad e_1e_2P_{\pm 12}=P_{\pm 12}e_1e_2, \quad eP_{\pm e}=P_{\pm e}e,
\end{equation}
\begin{equation}\label{3.8}
P_{\pm 0}=\pm P_{\pm 0}e_0,  \quad  P_{\pm 12}=\pm iP_{\pm 12}e_1e_2, \quad  P_{\pm e}=\pm iP_{\pm e}e.
\end{equation}
\end{prop}
\bigskip
\begin{proof}
The proof is a computation.
 \end{proof}
\bigskip
\begin{prop}
For any inhomogeneous form $\Omega\in K(4)$ we have
\begin{equation}\label{3.9}
(d^c+\delta^c)\Omega=\sum_{\mu=0}^3e_\mu\Delta_\mu\Omega,
\end{equation}
where $\Delta_\mu$ is the difference operator which acts on each component of $\Omega$ by the rule~(\ref{2.6}).
\end{prop}
\bigskip
\begin{proof}
See Proposition~1 in  \cite{S3}.
\end{proof}

Thus the discrete Dirac-K\"{a}hler equation can be rewritten in the form
\begin{equation*}
i\sum_{i=0}^3e_\mu\Delta_\mu\Omega=m\Omega.
\end{equation*}
Clearly, this is a discrete analogue of Equation~(\ref{1.2}).
Let $\Omega\in K^{ev}(4)$ be a real-valued even inhomogeneous form.  A discrete analogue of the Hestenes equation (\ref{1.3}) is defined by
\begin{equation}\label{3.10}
-(d^c+\delta^c)\Omega e_1e_2=m\Omega e_0,
\end{equation}
where $e_0, e_1, e_2$ are given by  (\ref{3.1}).
Using  (\ref{2.7})--(\ref{2.13}) this equation  can be expressed in terms of difference equations as
\begin{eqnarray*}\label{}
\Delta_0\omega_k^{12}-\Delta_1\omega_k^{02}+\Delta_2\omega_k^{01}+\Delta_3\overset{4}{\omega}_k=m\overset{0}{\omega}_k,\\
\Delta_2\overset{0}{\omega}_k+\Delta_0\omega_k^{02}-\Delta_1\omega_k^{12}+\Delta_3\omega_k^{23}=m\omega_k^{01},\\
-\Delta_1\overset{0}{\omega}_k-\Delta_0\omega_k^{01}-\Delta_2\omega_k^{12}-\Delta_3\omega_k^{13}=m\omega_k^{02},\\
-\Delta_1\omega_k^{23}+\Delta_2\omega_k^{13}-\Delta_3\omega_k^{12}-\Delta_0\overset{4}{\omega}_k=m\omega_k^{03},\\
-\Delta_0\overset{0}{\omega}_k-\Delta_1\omega_k^{01}-\Delta_2\omega_k^{02}-\Delta_3\omega_k^{03}=m\omega_k^{12},\\
-\Delta_0\omega_k^{23}+\Delta_2\omega_k^{03}-\Delta_3\omega_k^{02}-\Delta_1\overset{4}{\omega}_k=m\omega_k^{13},\\
\Delta_0\omega_k^{13}-\Delta_1\omega_k^{03}+\Delta_3\omega_k^{01}-\Delta_2\overset{4}{\omega}_k=m\omega_k^{23},\\
\Delta_3\overset{0}{\omega}_k+\Delta_0\omega_k^{03}-\Delta_1\omega_k^{13}-\Delta_2\omega_k^{23}=m\overset{4}{\omega}_k.
\end{eqnarray*}
Consider the equation
\begin{equation}\label{3.11}
i(d^c+\delta^c)\Omega=m\Omega e_0,
\end{equation}
where $\Omega\in K^{ev}(4)$ is a complex-valued even inhomogeneous form. This is a discrete analogue of the Joyce equation (\ref{1.4}).
Substituting (\ref{2.7})--(\ref{2.13}) into  (\ref{3.11}) yields the difference version of this equation
\begin{align*}\label{}
i(\Delta_0\overset{0}{\omega}_k+\Delta_1\omega_k^{01}+\Delta_2\omega_k^{02}+\Delta_3\omega_k^{03})=m\overset{0}{\omega}_k,\\
i(\Delta_1\overset{0}{\omega}_k+\Delta_0\omega_k^{01}+\Delta_2\omega_k^{12}+\Delta_3\omega_k^{13})=-m\omega_k^{01},\\
i(\Delta_2\overset{0}{\omega}_k+\Delta_0\omega_k^{02}-\Delta_1\omega_k^{12}+\Delta_3\omega_k^{23})=-m\omega_k^{02},\\
i(\Delta_3\overset{0}{\omega}_k+\Delta_0\omega_k^{03}-\Delta_1\omega_k^{13}-\Delta_2\omega_k^{23})=-m\omega_k^{03},\\
i(\Delta_0\omega_k^{12}-\Delta_1\omega_k^{02}+\Delta_2\omega_k^{01}+\Delta_3\overset{4}{\omega}_k)=m\omega_k^{12},\\
i(\Delta_0\omega_k^{13}-\Delta_1\omega_k^{03}+\Delta_3\omega_k^{01}-\Delta_2\overset{4}{\omega}_k)=m\omega_k^{13},\\
i(\Delta_0\omega_k^{23}-\Delta_2\omega_k^{03}+\Delta_3\omega_k^{02}+\Delta_1\overset{4}{\omega}_k)=m\omega_k^{23},\\
i(\Delta_1\omega_k^{23}-\Delta_2\omega_k^{13}+\Delta_3\omega_k^{12}+\Delta_0\overset{4}{\omega}_k)=-m\overset{4}{\omega}_k.
\end{align*}
First suppose that the discrete Hestenes equation (\ref{3.10}) and Equation~(\ref{3.11}) act in $K(4)$, i.e. act in the same space as  the discrete Dirac-K\"{a}hler equation.
\bigskip
 \begin{prop}
Let $\Omega\in K(4)$ be a solution of Equation~(\ref{3.11}),  then
\begin{equation}\label{3.12}
\Omega=\Omega P_{+0}+\Omega P_{-0},
\end{equation}
where
$\Omega P_{+0}$  satisfies the discrete Dirac-K\"{a}hler equation (\ref{2.14}) while
$\Omega P_{-0}$  satisfies the same equation but the sign of the right hand side changed to its opposite.
\end{prop}
\bigskip
\begin{proof}
By the definition of $P_{\pm 0}$  (\ref{3.5}) the decomposition (\ref{3.12}) is true for any form  $\Omega\in K(4)$.
Substituting $\Omega P_{\pm 0}$ into (\ref{3.11})   gives by (\ref{3.8})
\begin{equation*}
i(d^c+\delta^c)\Omega P_{\pm 0}=m\Omega e_0P_{\pm 0}=\pm m\Omega P_{\pm 0}.
\end{equation*}
\end{proof}
\bigskip
\begin{prop}
Let $\Omega\in K(4)$ be a solution of Equation~(\ref{3.11}),  then
\begin{equation}\label{3.13}
\Omega=\Omega P_{+12}+\Omega P_{-12},
\end{equation}
where
$\Omega P_{+12}$  satisfies the discrete Hestenes equation (\ref{3.10}) while
$\Omega P_{-12}$  satisfies the same equation but the sign of the right hand side changed to its opposite.
\end{prop}
\bigskip
\begin{proof} The proof is basically the same as for the previous proposition.
\end{proof}
\bigskip
\begin{cor}
Every solution $\Omega\in K(4)$ of Equation~(\ref{3.11}) decomposes into four parts
\begin{equation}\label{3.14}
\Omega=\Omega P_{+0}P_{+12}+\Omega P_{+0}P_{-12}+\Omega P_{-0}P_{+12}+\Omega P_{-0}P_{-12},
\end{equation}
where any of the parts  $\Omega P_{\pm 0}P_{\pm 12}$ satisfies both  the discrete Dirac-K\"{a}hler equation and  the discrete Hestenes equation with corresponding
sign of the right hand side.
\end{cor}
\bigskip
\begin{proof} By (\ref{3.12}) and (\ref{3.13}) any form $\Omega\in K(4)$ may be written as
\begin{equation*}
\Omega=(\Omega P_{+0}+\Omega P_{-0})P_{+12}+(\Omega P_{+0}+\Omega P_{-0})P_{-12}.
\end{equation*}
This gives (\ref{3.14}), and from Propositions~3.4 and 3.5 the claim follows.
\end{proof}

It should be noted that if  $\Omega$ is a solution of the discrete Hestenes equation then $\Omega e_2e_3 $ is a solution of the same equation with the sign of the right hand side reversed. Indeed by (\ref{3.2}), we have
\begin{equation*}
-(d^c+\delta^c)(\Omega  e_2e_3)e_1e_2=(d^c+\delta^c)\Omega  e_1e_2e_2e_3=-m\Omega e_0e_2e_3=-m(\Omega e_2e_3)e_0.
\end{equation*}
Hence the transformation $\Omega\rightarrow\Omega e_2e_3 $ changes the sign of the mass term in the discrete Hestenes equation.
Apply this transformation to (\ref{3.13}). Note that
\begin{equation*}
P_{\pm12}e_2e_3=e_2e_3P_{\mp12}.
\end{equation*}
It follows that the map $\Omega\rightarrow\Omega e_2e_3 $  transforms the part $\Omega P_{+12}$ into $\Omega P_{-12}$ and vice versa.
This gives the equivalence of both the parts in the decomposition (\ref{3.13}) (see \cite{JM} for the continuum case).

Now we return to our initial spaces of solutions of Equations~(\ref{3.10}) and (\ref{3.11}).
The point is to represent a complex-valued even solution of Equation~(\ref{3.11}) via  real-valued even solutions of  Equation~(\ref{3.10}).
Let  $\Omega\in K^{ev}(4)$ be a complex-valued even inhomogeneous form and $\Omega$ is  a solution of Equation~(\ref{3.11}). Let $\overline{\Omega}$  be the complex conjugate of $\Omega$.
 By analogy with the continuum case \cite{B} we consider the real-valued even forms $\Omega_+$  and $\Omega_-$ given by
\begin{equation}\label{3.15}
\Omega_{\pm}=\frac{1}{2}(\Omega+\overline{\Omega})\pm\frac{i}{2}(\Omega-\overline{\Omega})e_1e_2.
\end{equation}
By (\ref{3.7}) and (\ref{3.8}) it is easy to check that
\begin{equation*}
\Omega P_{\pm12}=\Omega_{\pm}P_{\pm12}.
\end{equation*}
Hence, the even  complex-valued forms  $\Omega_{+}P_{+12}$ and $\Omega_{-}P_{-12}$ are solutions of the discrete Hestenes equation Equation~(\ref{3.10}) with the correct or reversed sign on the right hand side.
Since the discrete Hestenes equation is real and linear,  the real and imaginary parts of these complex-valued forms are also solutions of Equation~(\ref{3.10}). The real and  imaginary parts of $\Omega_{\pm}P_{\pm12}$ are
\begin{equation*}
\RE(\Omega_{\pm}P_{\pm12})=\frac{1}{2}\Omega_{\pm}, \qquad
\IM(\Omega_{\pm}P_{\pm12})=\pm\frac{1}{2}\Omega_{\pm}e_1e_2.
\end{equation*}
Thus we have the following assertion.
\bigskip
\begin{prop} Every solution $\Omega\in K^{ev}(4)$ of Equation~(\ref{3.11}) takes the form
\begin{equation}\label{3.16}
\Omega=\frac{1}{2}(\Omega_++\Omega_-)+\frac{i}{2}(\Omega_+-\Omega_-)e_1e_2.
\end{equation}
\end{prop}
\bigskip
Each term of the real and imaginary parts of the representation (\ref{3.16}) is a real even solution of the discrete Hestenes equation with the correct or reversed sign on the right hand side. The formula  (\ref{3.16}) is a straightforward discrete version of the continuum counterpart (see \cite{B}).

Let us consider the following equation
\begin{equation}\label{3.17}
-(d^c+\delta^c)\Omega=m\Omega e,
\end{equation}
where   $\Omega\in K(4)$ is an arbitrary  real-valued
 inhomogeneous form and  $e$ is given by  (\ref{3.3}). This is a discrete analogue of Equation~(\ref{1.5}).
 \bigskip
\begin{prop}
Let $\Omega\in K(4)$ be a solution of Equation~(\ref{3.17}),  then
\begin{equation}\label{3.18}
\Omega=\Omega P_{-e}+\Omega P_{+e},
\end{equation}
where
$\Omega P_{-e}$  satisfies the discrete Dirac-K\"{a}hler equation (\ref{2.14}) while
$\Omega P_{+e}$  satisfies the same equation but the sign of the right hand side changed to its opposite.
\end{prop}
\bigskip
\begin{proof}
By use properties (\ref{3.7}) and (\ref{3.8}) of the projectors $P_{\pm e}$ one has
\begin{equation*}
i(d^c+\delta^c)\Omega P_{-e}=-im\Omega eP_{-e}=-im\Omega P_{-e}e=-i^2m\Omega P_{-e}=m\Omega P_{-e},
\end{equation*}
and
\begin{equation*}
i(d^c+\delta^c)\Omega P_{+e}=-im\Omega eP_{+e}=-im\Omega P_{+e}e=i^2m\Omega P_{+e}=-m\Omega P_{+e}.
\end{equation*}
\end{proof}

Consider now the transformation $\Omega\rightarrow\Omega e_1e_2e_3$. By (\ref{3.4})
it is easy to check that
\begin{equation*}
P_{\pm e}e_1e_2e_3=e_1e_2e_3P_{\mp e}.
\end{equation*}
From this on substituting $\Omega e_1e_2e_3$ in (\ref{3.18}) we obtain
\begin{equation*}
\Omega e_1e_2e_3=\Omega e_1e_2e_3 P_{-e}+\Omega e_1e_2e_3 P_{+e}=\Omega P_{+e}e_1e_2e_3+\Omega P_{-e}e_1e_2e_3.
\end{equation*}
Hence,  the map $\Omega\rightarrow\Omega e_1e_2e_3$ transforms the solution $\Omega P_{-e}$ of the discrete Dirac-K\"{a}hler equation into
the solution of the same equation with reversed sign of the mass term and vice versa. Again, as in the case of decomposition (\ref{3.13}), we have  the equivalence of
both the parts $\Omega P_{-e}$ and $\Omega P_{+e}$ in (\ref{3.18}).

Assume the discrete Hestenes equation   acts in  the larger space of complex-valued forms $K(4)$.
\bigskip
 \begin{prop}
Let $\Omega\in K(4)$ be a solution of Equation~(\ref{3.17}),  then
\begin{eqnarray*}\label{}
\Omega=\Omega P_{-e}P_{+0}P_{+12}+\Omega P_{+e}P_{+0}P_{+12}+\Omega P_{-e}P_{-0}P_{+12}+\Omega P_{+e}P_{-0}P_{+12}\\
+\Omega P_{-e} P_{+0}P_{-12}+\Omega P_{+e}P_{+0}P_{-12}+\Omega P_{-e}P_{-0}P_{-12}+\Omega P_{+e}P_{-0}P_{-12},
\end{eqnarray*}
where
$\Omega P_{-e}P_{+0}P_{+12}$, \ $\Omega P_{+e}P_{-0}P_{+12}$, \  $\Omega P_{+e}P_{+0}P_{-12}$ and \ $\Omega P_{-e}P_{-0}P_{-12}$ satisfy the discrete Hestenes equation (\ref{3.10}) while
$\Omega P_{+e}P_{+0}P_{+12}$, \ $\Omega P_{-e}P_{-0}P_{+12}$, \ $\Omega P_{-e}P_{+0}P_{-12}$ and \ $\Omega P_{+e}P_{-0}P_{-12}$  satisfy the same equation but the sign of the right hand side changed to its opposite.
\end{prop}
\bigskip
\begin{proof}
Substituting (\ref{3.18}) into (\ref{3.14}) yields the above representation for any form $\Omega\in K(4)$.
Let us first check that the part $\Omega P_{-e}P_{+0}P_{+12}$ satisfies the discrete Hestenes equation.
Since $\Omega$ is a solution of (\ref{3.17}), using (\ref{3.7}) and  (\ref{3.8}), we obtain
\begin{eqnarray*}
-(d^c+\delta^c)\Omega P_{-e}P_{+0}P_{+12}=m\Omega eP_{-e}P_{+0}P_{+12}=im\Omega P_{-e}P_{+0}P_{+12}\\
=im\Omega P_{-e}P_{+0}e_0P_{+12}=im\Omega P_{-e}P_{+0}P_{+12}e_0.
\end{eqnarray*}
On the other hand, we have
\begin{equation*}
-(d^c+\delta^c)\Omega P_{-e}P_{+0}P_{+12}=-i(d^c+\delta^c)\Omega P_{-e}P_{+0}P_{+12}e_1e_2.
\end{equation*}
From the above it follows that
\begin{equation*}
-(d^c+\delta^c)(\Omega P_{-e}P_{+0}P_{+12})e_1e_2=m(\Omega P_{-e}P_{+0}P_{+12})e_0.
\end{equation*}
The same proof remains valid for all other cases.
\end{proof}

Now our aim is to express a solution of the discrete Dirac-K\"{a}hler equation (\ref{2.14}) via solutions of Equation~(\ref{3.17}) which are real-valued.
Let  $\Omega\in K(4)$ be a complex-valued solution of  Equation~(\ref{2.14}).
 Consider the following real-valued forms
 \begin{equation}\label{3.19}
\Omega_{\pm}=\frac{1}{2}(\Omega+\overline{\Omega})\pm\frac{i}{2}(\Omega-\overline{\Omega})e,
\end{equation}
where $\overline{\Omega}$  is the complex conjugate of $\Omega$ and  $e$ is given by  (\ref{3.3}).
\bigskip
\begin{prop} Every solution $\Omega\in K(4)$ of the discrete Dirac-K\"{a}hler equation (\ref{2.14}) takes the form:
\begin{equation}\label{3.20}
\Omega=\frac{1}{2}(\Omega_++\Omega_-)+\frac{i}{2}(\Omega_+-\Omega_-)e,
\end{equation}
where $\frac{1}{2}\Omega_-$ and $\frac{1}{2}\Omega_-e$ are the solutions of Equation~(\ref{3.17}) while  $\frac{1}{2}\Omega_+$ and $\frac{1}{2}\Omega_+e$ are the solutions
of the same equations with the reversed sign of the mass term.
\end{prop}
\bigskip
\begin{proof}
The decomposition (\ref{3.18}) holds for any form. If  $\Omega\in K(4)$ is a solution of Equation~(\ref{2.14}), then $\Omega P_{-e}$ is a solution of Equation~(\ref{3.17}) and  $\Omega P_{+e}$ satisfies Equation~(\ref{3.17}) with the opposite sign of the mass term. Since
\begin{equation*}
\Omega_- P_{-e}=\Omega P_{-e}, \qquad \Omega_+ P_{+e}=\Omega P_{+e},
\end{equation*}
then $\Omega_{\mp} P_{\mp e}$ are also solutions of  Equation~(\ref{3.17}) with  the correct or opposite sign of the mass term correspondingly. Note that  $\Omega_{\mp} P_{\mp e}$ are complex-valued.
Using (\ref{3.5})  one easily checks that the real and imaginary parts of these complex-valued forms are
\begin{equation*}
\RE(\Omega_{\mp}P_{\mp e})=\frac{1}{2}\Omega_{\mp}, \qquad
\IM(\Omega_{\mp}P_{\mp e})=\mp\frac{1}{2}\Omega_{\mp}e.
\end{equation*}
Since Equation~(\ref{3.17}) is real and linear, then the real and imaginary parts of a solution of with equation  are also solutions. Thus $\frac{1}{2}\Omega_{\mp}$ and $\frac{1}{2}\Omega_{\mp}e$ are real-valued solutions of Equation~(\ref{3.17}) with the corresponding sign of  the mass term.
\end{proof}


\begin{thebibliography}{99}
\bibitem[1]{B}
\textsc{W. E. Baylis.} Comment on "Dirac theory in spacetime algebra".
\textit{J. Phys. A: Math. Gen.} \textbf{35} (2002), 4791--4796.

\bibitem[2]{Dezin}
\textsc{A. A. Dezin.}
Multidimensional Analysis and Discrete Models.
CRC Press, Boca Raton, 1995.

\bibitem[3]{H1}
\textsc{D. Hestenes.} Real Spinor Fields.
\textit{Journal of Mathematical Physics.} \textbf{8}, \textsl{4} (1967), 798--808.

\bibitem[4]{H2} \textsc{D. Hestenes.} Spacetime Algebra. Gordon and Breach,  New York, 1966.

\bibitem[5]{J}
\textsc{W. P. Joyce.} Dirac theory in spacetime algebra: I. The generalized
bivector Dirac equation.
\textit{J. Phys. A: Math. Gen.} \textbf{34} (2001), 1991--2005.

\bibitem[6]{JM}
\textsc{W. P. Joyce, J. G. Martin.} Equivalence of Dirac formulations.
\textit{J. Phys. A: Math. Gen.} \textbf{35}  (2002), 4729--4736.

\bibitem[7]{Kahler}
 \textsc{E. K\"{a}hler.}  Der innere differentialk\"{u}l. \textit{Rendiconti di Matematica.}  \textbf{21}, \textsl{3--4} (1962), 425--523.

\bibitem[8]{Marchuk}
\textsc{N. G.  Marchuk.}   Dirac-type tensor equations. \textit{Nuovo Cimento Soc. Ital. Fis. B.} \textbf{116},  \textsl{11} (2001), 1225--1248.

\bibitem[9]{Rabin}
\textsc{J. M. Rabin.} Homology theory of lattice fermion doubling. \textit{Nucl. Phys. B.}  \textbf{201}, \textsl{2} (1982),  315--332.

\bibitem[10]{S1}
\textsc{V. Sushch.} A discrete model of the Dirac-K\"{a}hler equation. \textit{Rep. Math. Phys.} \textbf{73}, \textsl{1} (2014),  109--125. arXiv:1307.1220v2.

\bibitem[11]{S2}
\textsc{V. Sushch.} On the chirality of a discrete Dirac-K\"{a}hler equation.  \textit{Rep. Math. Phys.} \textbf{76}, \textsl{2} (2015), 179--196. arXiv:1411.7673.

\bibitem[12]{S3}
\textsc{V. Sushch.} Discrete Dirac-K\"{a}hler and Hestenes equations. Differential and difference equations with applications.
Springer Proc. Math. Stat. \textbf{164}, 433--442.  Springer, New York, 2016. arXiv:1509.06907v1.


\end{thebibliography}
\end{document}